\documentclass[aps,prl,superscriptaddress,showpacs,floatfix,twocolumn]{revtex4}

\usepackage{graphicx}   

\usepackage{dcolumn}  

\usepackage{xspace} 

\newcommand{\pt}{\mbox{$p_{\rm T}$}\xspace}
\newcommand{\punit}{\mbox{GeV$/c$}\xspace}
\newcommand{\munit}{\mbox{MeV$/c^2$}\xspace}

\newcommand{\all}{\mbox{$A_{\rm LL}$}\xspace}
\newcommand{\allpz}{\mbox{$A_{\rm LL}^{\pi^0}$}\xspace}
\newcommand{\dg}{\mbox{$\Delta G$}\xspace}
\newcommand{\dgx}{\mbox{$\Delta g(x)$}\xspace}
\newcommand{\tdg}{\mbox{$\Delta G^{[0.02,0.3]}$}\xspace}
\newcommand{\pz}{\mbox{$\pi^0$}\xspace}
\newcommand{\chisq}{\mbox{$\chi^2$}\xspace}

\begin{document}


\title{The gluon spin contribution to the proton spin from the 
double helicity asymmetry in inclusive $\pi^{0}$ production in 
polarized $p + p$ collisions at $\sqrt{s} = 200$ GeV}

\newcommand{\abilene}{Abilene Christian University, Abilene, TX 79699, U.S.}
\newcommand{\acadsin}{Institute of Physics, Academia Sinica, Taipei 11529, Taiwan}
\newcommand{\banaras}{Department of Physics, Banaras Hindu University, Varanasi 221005, India}
\newcommand{\barc}{Bhabha Atomic Research Centre, Bombay 400 085, India}
\newcommand{\bnlchem}{Chemistry Department, Brookhaven National Laboratory, Upton, NY 11973-5000, U.S.}
\newcommand{\bnlcoll}{Collider-Accelerator Department, Brookhaven National Laboratory, Upton, NY 11973-5000, U.S.}
\newcommand{\bnlphys}{Physics Department, Brookhaven National Laboratory, Upton, NY 11973-5000, U.S.}
\newcommand{\caucr}{University of California - Riverside, Riverside, CA 92521, U.S.}
\newcommand{\charlesczech}{Charles University, Ovocn\'{y} trh 5, Praha 1, 116 36, Prague, Czech Republic}
\newcommand{\ciae}{China Institute of Atomic Energy (CIAE), Beijing, People's Republic of China}
\newcommand{\cns}{Center for Nuclear Study, Graduate School of Science, University of Tokyo, 7-3-1 Hongo, Bunkyo, Tokyo 113-0033, Japan}
\newcommand{\colorado}{University of Colorado, Boulder, CO 80309, U.S.}
\newcommand{\columbia}{Columbia University, New York, NY 10027 and Nevis Laboratories, Irvington, NY 10533, U.S.}
\newcommand{\czechtech}{Czech Technical University, Zikova 4, 166 36 Prague 6, Czech Republic}
\newcommand{\dapnia}{Dapnia, CEA Saclay, F-91191, Gif-sur-Yvette, France}
\newcommand{\debrecen}{Debrecen University, H-4010 Debrecen, Egyetem t{\'e}r 1, Hungary}
\newcommand{\elte}{ELTE, E{\"o}tv{\"o}s Lor{\'a}nd University, H - 1117 Budapest, P{\'a}zm{\'a}ny P. s. 1/A, Hungary}
\newcommand{\fit}{Florida Institute of Technology, Melbourne, FL 32901, U.S.}
\newcommand{\fsu}{Florida State University, Tallahassee, FL 32306, U.S.}
\newcommand{\gsu}{Georgia State University, Atlanta, GA 30303, U.S.}
\newcommand{\hiroshima}{Hiroshima University, Kagamiyama, Higashi-Hiroshima 739-8526, Japan}
\newcommand{\ihepprot}{IHEP Protvino, State Research Center of Russian Federation, Institute for High Energy Physics, Protvino, 142281, Russia}
\newcommand{\illuiuc}{University of Illinois at Urbana-Champaign, Urbana, IL 61801, U.S.}
\newcommand{\instpasczech}{Institute of Physics, Academy of Sciences of the Czech Republic, Na Slovance 2, 182 21 Prague 8, Czech Republic}
\newcommand{\isu}{Iowa State University, Ames, IA 50011, U.S.}
\newcommand{\jinrdubna}{Joint Institute for Nuclear Research, 141980 Dubna, Moscow Region, Russia}
\newcommand{\kek}{KEK, High Energy Accelerator Research Organization, Tsukuba, Ibaraki 305-0801, Japan}
\newcommand{\kfki}{KFKI Research Institute for Particle and Nuclear Physics of the Hungarian Academy of Sciences (MTA KFKI RMKI), H-1525 Budapest 114, POBox 49, Budapest, Hungary}
\newcommand{\korea}{Korea University, Seoul, 136-701, Korea}
\newcommand{\kurchatov}{Russian Research Center ``Kurchatov Institute", Moscow, Russia}
\newcommand{\kyoto}{Kyoto University, Kyoto 606-8502, Japan}
\newcommand{\labllr}{Laboratoire Leprince-Ringuet, Ecole Polytechnique, CNRS-IN2P3, Route de Saclay, F-91128, Palaiseau, France}
\newcommand{\lawllnl}{Lawrence Livermore National Laboratory, Livermore, CA 94550, U.S.}
\newcommand{\losalamos}{Los Alamos National Laboratory, Los Alamos, NM 87545, U.S.}
\newcommand{\lpc}{LPC, Universit{\'e} Blaise Pascal, CNRS-IN2P3, Clermont-Fd, 63177 Aubiere Cedex, France}
\newcommand{\lund}{Department of Physics, Lund University, Box 118, SE-221 00 Lund, Sweden}
\newcommand{\mass}{Department of Physics, University of Massachusetts, Amherst, MA 01003-9337, U.S. }
\newcommand{\muenster}{Institut f\"ur Kernphysik, University of Muenster, D-48149 Muenster, Germany}
\newcommand{\muhlenberg}{Muhlenberg University, Allentown, PA 18104-5586, U.S.}
\newcommand{\myongji}{Myongji University, Yongin, Kyonggido 449-728, Korea}
\newcommand{\nagasaki}{Nagasaki Institute of Applied Science, Nagasaki-shi, Nagasaki 851-0193, Japan}
\newcommand{\newmex}{University of New Mexico, Albuquerque, NM 87131, U.S. }
\newcommand{\nmsu}{New Mexico State University, Las Cruces, NM 88003, U.S.}
\newcommand{\ornl}{Oak Ridge National Laboratory, Oak Ridge, TN 37831, U.S.}
\newcommand{\orsay}{IPN-Orsay, Universite Paris Sud, CNRS-IN2P3, BP1, F-91406, Orsay, France}
\newcommand{\peking}{Peking University, Beijing, People's Republic of China}
\newcommand{\pnpi}{PNPI, Petersburg Nuclear Physics Institute, Gatchina, Leningrad region, 188300, Russia}
\newcommand{\riken}{RIKEN, The Institute of Physical and Chemical Research, Wako, Saitama 351-0198, Japan}
\newcommand{\rikjrbrc}{RIKEN BNL Research Center, Brookhaven National Laboratory, Upton, NY 11973-5000, U.S.}
\newcommand{\rikkyo}{Physics Department, Rikkyo University, 3-34-1 Nishi-Ikebukuro, Toshima, Tokyo 171-8501, Japan}
\newcommand{\saispbstu}{Saint Petersburg State Polytechnic University, St. Petersburg, Russia}
\newcommand{\saopaulo}{Universidade de S{\~a}o Paulo, Instituto de F\'{\i}sica, Caixa Postal 66318, S{\~a}o Paulo CEP05315-970, Brazil}
\newcommand{\seoulnat}{System Electronics Laboratory, Seoul National University, Seoul, Korea}
\newcommand{\stonybrkc}{Chemistry Department, Stony Brook University, Stony Brook, SUNY, NY 11794-3400, U.S.}
\newcommand{\stonycrkp}{Department of Physics and Astronomy, Stony Brook University, SUNY, Stony Brook, NY 11794, U.S.}
\newcommand{\subatech}{SUBATECH (Ecole des Mines de Nantes, CNRS-IN2P3, Universit{\'e} de Nantes) BP 20722 - 44307, Nantes, France}
\newcommand{\tenn}{University of Tennessee, Knoxville, TN 37996, U.S.}
\newcommand{\titech}{Department of Physics, Tokyo Institute of Technology, Oh-okayama, Meguro, Tokyo 152-8551, Japan}
\newcommand{\tsukuba}{Institute of Physics, University of Tsukuba, Tsukuba, Ibaraki 305, Japan}
\newcommand{\vandy}{Vanderbilt University, Nashville, TN 37235, U.S.}
\newcommand{\waseda}{Waseda University, Advanced Research Institute for Science and Engineering, 17 Kikui-cho, Shinjuku-ku, Tokyo 162-0044, Japan}
\newcommand{\weizmann}{Weizmann Institute, Rehovot 76100, Israel}
\newcommand{\yonsei}{Yonsei University, IPAP, Seoul 120-749, Korea}
\affiliation{\abilene}
\affiliation{\acadsin}
\affiliation{\banaras}
\affiliation{\barc}
\affiliation{\bnlchem}
\affiliation{\bnlcoll}
\affiliation{\bnlphys}
\affiliation{\caucr}
\affiliation{\charlesczech}
\affiliation{\ciae}
\affiliation{\cns}
\affiliation{\colorado}
\affiliation{\columbia}
\affiliation{\czechtech}
\affiliation{\dapnia}
\affiliation{\debrecen}
\affiliation{\elte}
\affiliation{\fit}
\affiliation{\fsu}
\affiliation{\gsu}
\affiliation{\hiroshima}
\affiliation{\ihepprot}
\affiliation{\illuiuc}
\affiliation{\instpasczech}
\affiliation{\isu}
\affiliation{\jinrdubna}
\affiliation{\kek}
\affiliation{\kfki}
\affiliation{\korea}
\affiliation{\kurchatov}
\affiliation{\kyoto}
\affiliation{\labllr}
\affiliation{\lawllnl}
\affiliation{\losalamos}
\affiliation{\lpc}
\affiliation{\lund}
\affiliation{\mass}
\affiliation{\muenster}
\affiliation{\muhlenberg}
\affiliation{\myongji}
\affiliation{\nagasaki}
\affiliation{\newmex}
\affiliation{\nmsu}
\affiliation{\ornl}
\affiliation{\orsay}
\affiliation{\peking}
\affiliation{\pnpi}
\affiliation{\riken}
\affiliation{\rikjrbrc}
\affiliation{\rikkyo}
\affiliation{\saispbstu}
\affiliation{\saopaulo}
\affiliation{\seoulnat}
\affiliation{\stonybrkc}
\affiliation{\stonycrkp}
\affiliation{\subatech}
\affiliation{\tenn}
\affiliation{\titech}
\affiliation{\tsukuba}
\affiliation{\vandy}
\affiliation{\waseda}
\affiliation{\weizmann}
\affiliation{\yonsei}
\author{A.~Adare}	\affiliation{\colorado}
\author{S.~Afanasiev}	\affiliation{\jinrdubna}
\author{C.~Aidala}	\affiliation{\mass}
\author{N.N.~Ajitanand}	\affiliation{\stonybrkc}
\author{Y.~Akiba}	\affiliation{\riken} \affiliation{\rikjrbrc}
\author{H.~Al-Bataineh}	\affiliation{\nmsu}
\author{J.~Alexander}	\affiliation{\stonybrkc}
\author{K.~Aoki}	\affiliation{\kyoto} \affiliation{\riken}
\author{L.~Aphecetche}	\affiliation{\subatech}
\author{J.~Asai}	\affiliation{\riken}
% ...
\collaboration{PHENIX Collaboration} \noaffiliation

\date{\today}

\begin{abstract}

The double helicity asymmetry in neutral pion production for 
\pt=1 to 12 \punit has been measured with the PHENIX experiment in 
order to access the gluon spin contribution, \dg, to the proton spin.  
Measured asymmetries are consistent with zero, and at a theory scale 
of $\mu^2=4$ GeV$^2$ give $\tdg=0.1$ to 0.2, with a constraint of 
$-0.7<\tdg<0.5$ at $\Delta \chisq=9$ ($\sim$3$\sigma$) for our sampled gluon 
momentum fraction ($x$) range, 0.02 to 0.3.  The results are obtained 
using predictions for our measured asymmetries generated from four 
representative fits to polarized deep inelastic scattering data.  We 
also consider the dependence of the \dg constraint on the choice of 
theoretical scale, a dominant uncertainty in these predictions.

\end{abstract}

\pacs{13.85.Ni,13.88.+e,21.10.Hw,25.40.Ep}

\maketitle


The quark spin contribution to the proton spin was found to be 
only $\sim$25\% 
\cite{r:EMC:spincrisis2,r:Hermes:Airapetian:2007mh,r:COMPASS:Alexakhin:2006vx}, 
indicating that the majority of the proton spin on average 
comes from the gluon spin contribution, \dg, and/or from gluon 
and quark orbital angular momentum. High energy polarized 
proton-proton collisions at the Relativistic Heavy Ion 
Collider (RHIC) at Brookhaven National Laboratory access \dg 
at leading order through spin-dependent gluon-gluon 
(\textit{gg}) and quark-gluon (\textit{qg}) scattering.

This paper presents results from the 2006 RHIC run (Run-6) on 
\dg from measurements of the double helicity asymmetry (\all) 
in inclusive mid-rapidity \pz production by the PHENIX 
experiment. \dg can be extracted from \allpz using next to 
leading order (NLO) perturbative quantum chromodynamics (pQCD) 
\cite{r:DG-to-ALL:Jager:2002xm}, which successfully describes 
unpolarized cross-sections measured at RHIC for many inclusive 
processes 
\cite{r:STAR:SSApi0,r:STAR:jetcrosssection:Abelev:2006uq,r:PHENIX:directphoton_crosssection},
including mid-rapidity \pz production 
\cite{r:PHENIX:run5pi0all:Adare:2007dg}, at $\sqrt{s}$=200 
GeV. These data represent a factor of two improvement in the 
statistical uncertainty compared to previous results 
\cite{r:PHENIX:run3pi0,r:PHENIX:run4pi0,r:PHENIX:run5pi0all:Adare:2007dg}. 
They significantly constrain \dg, as presented in a recent 
global fit (DSSV) \cite{r:DSSV:deFlorian:2008mr} of both RHIC 
and polarized deep inelastic scattering (pDIS) data, which 
used a preliminary version of these results.  We further 
present the impact of experimental systematic and several 
theoretical uncertainties on our determination of \dg.

\begin{figure}[b]
\includegraphics[width=1.0\linewidth]{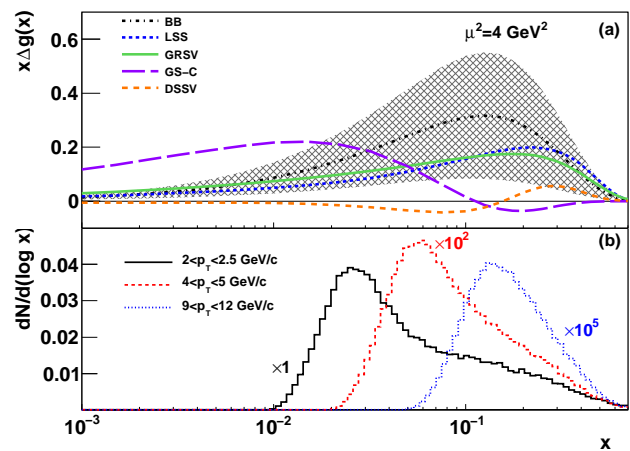}
\caption{\label{f:dg} (color online) (a)  
The polarized gluon distribution as a function of $x$ for five 
fits to polarized data.  Hatched band is pDIS $1\sigma$ 
uncertainty (BB).  (b) Distributions of gluon $x$ in three \pz 
\pt bins from a NLO pQCD simulation.}
\end{figure}

\begin{figure*}[hbtp]
\includegraphics[width=1.0\linewidth]{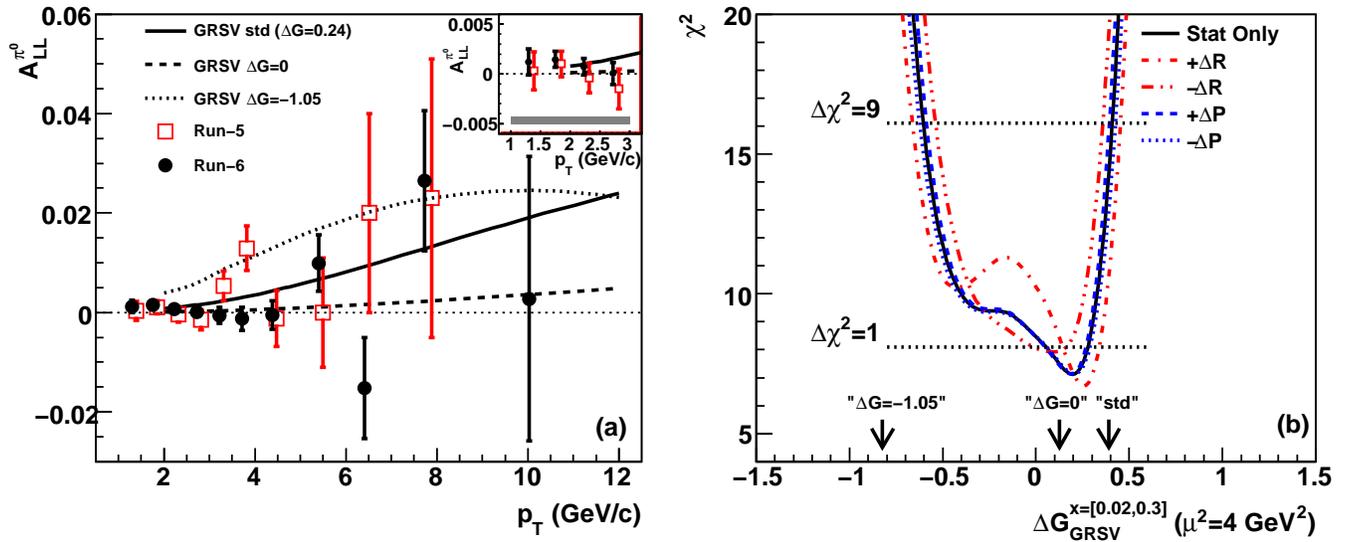}
\caption{\label{f:data} (color online) 
(a) Asymmetry in \pz production as a function of \pt.  Error 
bars are statistical uncertainties. An $8.3\%$ scale 
uncertainty due to the uncertainty in beam polarization is not 
shown. The \pt independent uncertainty of $7\times10^{-4}$ due 
to relative luminosity is shown only in the inset as a shaded 
bar. For comparison, we also show our Run-5 result.  NLO pQCD 
expectations based on different inputs for \dg in the GRSV 
parameterization are plotted. (b)  The 
\chisq profile as a function of $\dg^{[0.02,0.3]}_{\rm GRSV}$ 
using the combined Run5 and Run6 data considering only 
statistical uncertainty, or also varying by $\pm 1 \sigma$ the 
two primary experimental systematic uncertainties, from beam 
polarizations $(\pm \Delta P)$ and relative luminosity $(\pm 
\Delta R)$.}
\end{figure*}

We define $\dg^{[a,b]}\equiv\int_a^b dx\Delta g(x,\mu^2)$, 
with $\Delta g(x,\mu^2)$, the polarized gluon distribution, a 
function of $x$, the gluon momentum fraction and $\mu^2$, the 
factorization scale.  Thus $\dg^{[0,1]}\equiv\dg$. 
Figure~\ref{f:dg}a shows the best fit \dgx from four different 
pDIS fits: GRSV-std \cite{r:GRSVfit:Gluck:2000dy}, BB 
(``ISET-4'') \cite{r:BB2002:Bluemlein:2002be}, LSS 
\cite{r:LSS2006:Leader:2005ci} and GS-C 
\cite{r:GS1996:Gehrmann:1995ag} which assumes a node, or sign 
change, in \dgx. As the pDIS data have limited sensitivity to 
\dg, there remains large uncertainty, which was estimated by 
BB, and is shown as the hatched band.  The result from the 
recent global fit, DSSV, is also shown. It has a node, which 
is driven by the inclusion of RHIC PHENIX \pz and STAR jet 
\cite{r:STAR:run5jetall:Abelev:2007vt, r:STAR:run6jet} \all 
data along with evolution from pDIS at large $x$ 
\cite{r:DSSV:deFlorian:2008mr}.  Table~\ref{t:dgvalues} lists 
$\dg^{[a,b]}$ for several $x$ ranges. 

We define $\allpz = (\sigma_{++} - \sigma_{+-})/(\sigma_{++} + 
\sigma_{+-})$, with $\sigma_{++}$ ($\sigma_{+-}$) the beam 
helicity dependent differential cross sections for inclusive 
\pz production from collisions of longitudinally polarized 
protons with the same (opposite) helicity.  The asymmetry is 
measured using
\begin{equation}
 \allpz = \frac{1}{\langle P_{B}P_{Y}\rangle}\frac{N_{++} - R N_{+-}}{N_{++} + R N_{+-}},\;\;
    R = \frac{L_{++}}{L_{+-}}
 \label{e:all}
\end{equation}
where $P_{\rm B}$ and $P_{\rm Y}$ are the polarizations of the 
two RHIC beams, called ``Blue'' and ``Yellow,'' and $R$, the 
relative luminosity, is the ratio of integrated luminosities 
($L$) for same and opposite helicity collisions.  Here we take 
$N$ to be the \pz yield in a transverse momentum (\pt) bin.

Each \pz \pt bin samples a distribution in gluon $x$.  
Figure~\ref{f:dg}b shows the sampled gluon $x$ distributions 
for three \allpz \pt bins from a NLO pQCD simulation 
\cite{r:DG-to-ALL:Jager:2002xm,r:StratmanPrivateComm:2007}.  
They are peaked at $x_{\rm T}/0.7$ \cite{r:Adler:2006sc}, with 
$x_{\rm T}\equiv\pt/(\sqrt{s}/2)$.  The bins overlap, with our 
data covering primarily the range $0.02<x<0.3$, and so we 
probe \tdg.

The highly segmented PHENIX electromagnetic calorimeter 
(EMCal) \cite{r:PHENIX:EMCAL} is used to detect $\pz 
\rightarrow \gamma \gamma$ decays.  The EMCal covers a 
pseudorapidity range of $|\eta| < 0.35$ and azimuthal angle 
range of $ \Delta \phi =\pi$, with segmentation $\Delta \eta 
\times \Delta \phi=0.01 \times 0.01$.  We required for each of 
the two decay photons an energy deposition pattern consistent 
with an electromagnetic shower, no charged track pointing to 
the location of the deposited energy, and standard quality 
assurance requirements \cite{r:PHENIX:run3pi0}.  Events were 
obtained from an EMCal based high \pt photon trigger 
\cite{r:PHENIX:pi0_crosssection:Adler:2003pb} in coincidence 
with a minimum bias trigger 
\cite{r:PHENIX:run5pi0all:Adare:2007dg} (also used to obtain 
the relative luminosity).  This EMCal based trigger had an 
efficiency for \pz of $5\%$ at $\pt \approx 1$~\punit and 
$90\%$ for $\pt> 3.5$ \punit.  The minimum bias trigger was 
defined as the coincidence of signals from forward and 
backward beam-beam counters (BBC) with full azimuthal coverage 
located at pseudorapidities $\pm (3.0-3.9)$ 
\cite{r:PHENIX:BBC_NTC_MVD}. The analyzed data sample 
corresponded to an integrated luminosity of 6.5 pb$^{-1}$.

\begin{table}[b]

\caption{\dg for different $x$ ranges at $\mu^2=4$~GeV$^2$ for each 
group's best fit and the \chisq when comparing the expected 
\all in Fig. \ref{f:model}(a) with our data (8 degrees of 
freedom). Also, the minimum \chisq and corresponding \tdg 
found in Fig. \ref{f:model}(b).}

\label{t:dgvalues}
\begin{ruledtabular} \begin{tabular}{cccccc}
            & \multicolumn{3}{c}{Published best fit}
            & \multicolumn{2}{c}{From Fig. \ref{f:model}b} \\
 Group & $\dg^{[0,1]}$  & $\dg^{[0.02,0.3]}$ 
 & \chisq & $\dg^{[0.02,0.3]}$  & \chisq\\ \hline
  GS-C     &  0.95    &  0.18     &    8.3    & 0.1 &  8.5 \\
  DSSV    &  -0.05   &  -0.03    &   7.5     & NA & NA \\
  LSS       &  0.60     &  0.37     &   22.4  & 0.2 & 7.0 \\
  GRSV    &  0.67     &  0.38     &  14.8  & 0.2 & 7.1 \\
  BB          &  0.93     &  0.67     &  69.0  & 0.2 & 7.2 \\ 
\end{tabular} \end{ruledtabular}
\end{table}

Each collider ring of RHIC was filled with up to 111 out of a 
possible 120 bunches, spaced 106 ns apart, with bunch 
helicities set such that all four beam helicity combinations 
occurred in sequences of four bunch crossings.  The pattern of 
helicity combinations for each RHIC fill (typically 8 hrs) was 
cycled between four possibilities in order to reduce 
systematic uncertainties that could be correlated to the bunch 
structure in RHIC \cite{r:PHENIX:run5pi0all:Adare:2007dg}. 
Events were tagged with the bunch crossing number to obtain 
the beam helicities for the event. The luminosity weighted 
beam polarization product was $\langle P_{\rm B} P_{\rm 
Y}\rangle=0.322 \pm 0.027$ (8.3\%), with single beam 
polarizations of 0.560 and 0.575.  Using very forward neutron 
production asymmetry 
\cite{r:IP12:Bazilevsky:2006vd,r:PHENIX:run5pi0all:Adare:2007dg}, 
the longitudinal polarization fractions were found to be 
greater than 99\%.

As in our previous analyses 
\cite{r:PHENIX:run3pi0,r:PHENIX:run5pi0all:Adare:2007dg}, the 
relative luminosity ratio $R$ was obtained from 
crossing-by-crossing collected minimum bias (BBC) trigger 
counts, which measure about half of the $p+p$ inelastic cross 
section \cite{r:PHENIX:pi0_crosssection:Adler:2003pb}. The 
uncertainty on $R$ was derived from the comparison with a 
second trigger based on the Zero Degree Calorimeters 
\cite{r:RHIC:ZDC}, which selects different physics processes 
in a different kinematic range. It contributed a \pt 
independent systematic uncertainty to \all of 
$7\times10^{-4}$.

Equation~\ref{e:all} is used to determine, on a fill by fill 
basis, \all for the yield in the \pz mass peak for each \pt 
bin. The asymmetries were averaged over fills and corrected 
for the asymmetry in the background contribution (determined 
from two 50~\munit wide sidebands on either side of the \pz 
peak) \cite{r:PHENIX:run3pi0}, which was consistent with zero.

Figure~\ref{f:data}a shows the measured \allpz from Run-6 
\cite{r:data} in comparison with our published data from the 
2005 RHIC run (Run-5) \cite{r:PHENIX:run5pi0all:Adare:2007dg}.  
The results are found to be statistically consistent with a 
13\% confidence level.  The inset shows an expanded view of 
the low \pt region, as well as the relative luminosity 
uncertainty from Run-6. Besides this and the scale uncertainty 
from polarization, other systematic uncertainties that can be 
found by using a bunch polarization sign randomization 
technique and by varying the \pz identification criteria 
\cite{r:PHENIX:run3pi0} appear negligible.

Also shown in Fig.~\ref{f:data}a are NLO pQCD predictions of 
\allpz \cite{r:DG-to-ALL:Jager:2002xm} based on fits of pDIS 
data by GRSV with three different values for \dg at the {\it input} 
scale of $\mu^2=0.4$ GeV$^2$: 1) ``std'', their best fit value 
with $\dg=0.24$, 2)  $\dg=0$ and 3) $\dg=-1.05$.  The 
measurements are most consistent with GRSV $\dg=0$.  CTEQ6 
unpolarized parton distribution functions (PDF) 
\cite{r:CTEQ6:Pumplin:2002vw} were used, along with DSS 
fragmentation functions (FF) 
\cite{r:DSS2007:deFlorian:2007aj}, in all calculations.  
Using alternative PDF \cite{r:MRST:Martin:2002aw} or FF 
\cite{r:KKP:Kniehl:2000fe} did not lead to significant 
differences in the \all expectations.

\all expectations based on fits to the pDIS data with a range 
of inputs for $\dg^{[0,1]}$ evolved to $\mu^2=0.4$ GeV$^2$ in the GRSV 
parameterization were calculated.  Similar to our previous 
analysis \cite{r:PHENIX:run5pi0all:Adare:2007dg}, \chisq 
values were calculated using our combined Run-5 and Run-6 data 
\cite{r:data} for these expectations, effectively fitting \dg 
with our data in this parameterization. In Fig.~\ref{f:data}b, 
these values are plotted as a function of \tdg at 
$\mu^2=4$~GeV$^2$ . Due to soft physics contamination at low 
\pt, we use only data with $\pt>2$~\punit 
\cite{r:PHENIX:run5pi0all:Adare:2007dg}.  Assuming that 
$\mu=\pt$, $\mu^2=4$~GeV$^2$ is then the minimum cutoff of our 
data. The solid curve shows the result considering only 
statistical uncertainties. Due to \textit{gg} interactions in 
$p+p$ collisions, \all probes \dg quadratically in \pz 
production \cite{r:DG-to-ALL:Jager:2002xm}. The \chisq profile 
is thus not parabolic, and so we show 
$\Delta\chisq\equiv\chisq-\chi^2_{\rm min}=1$ and $9$ 
corresponding to ``$1\sigma$'' and ``$3\sigma$'' 
uncertainties.  The \textit{gg} scattering increases toward 
low \pt, which dominates our statistics and causes the two 
minima seen in Fig.~\ref{f:data}b.  The larger allowed 
negative region arises from cancelation of \textit{gg} and 
\textit{qg} terms in \allpz when \dg is negative.

For a robust interpretation of our results in terms of \dg, we 
consider not only statistical but also experimental systematic 
and theoretical uncertainties. The effects of the two largest 
experimental systematic uncertainties, due to polarization and 
relative luminosity, are shown in Fig.~\ref{f:data}b. The 
polarization uncertainty is insignificant when extracting \dg. 
However, the uncertainty on relative luminosity, though small, 
cannot be neglected.   
Accounting for statistical uncertainty, we find $\Delta
G^{[0.02,0.3]}_{\rm GRSV}=0.2\pm0.1$ ($1\sigma$) and
$0.2^{+0.2}_{-0.8}$ ($3\sigma$) with an additional experimental
systematic uncertainty of $\pm0.1$.

\begin{figure}[btp]
\includegraphics[width=1.0\linewidth]{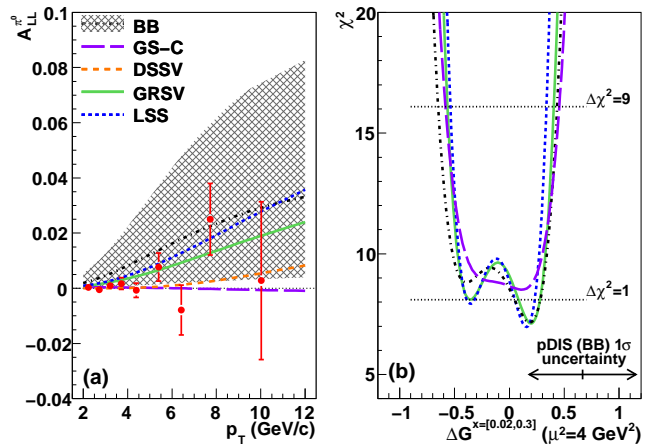}
\caption{\label{f:model} (color online)
(a) \pz asymmetry expectations for different \dgx in 
Fig.~\ref{f:dg}b.  Hatched band is the pDIS uncertainty (BB).  
Combined Run-5 and Run-6 results are also plotted (statistical 
errors only). (b) The \chisq profile as a function of \tdg for 
same parameterizations.  Arrows indicate $1\sigma$ uncertainty 
on BB best fit. $\Delta\chisq$ values are shown for GRSV.}
\end{figure}

Figure~\ref{f:model}a shows \all expectations 
\cite{r:DG-to-ALL:Jager:2002xm,r:StratmanPrivateComm:2007} 
based on the parameterizations discussed above, along with the 
pDIS uncertainty on \dgx in BB propagated to \all.  Our 
combined Run-5 and Run-6 \pz results \cite{r:data} are also 
plotted for comparison.  The \chisq values for comparing each 
curve with our data are given in Table~\ref{t:dgvalues}.  The 
three fit results without a node in \dgx--LSS, GRSV and 
BB--have large values of \tdg which lead to relatively large 
asymmetries that lie mostly above the data, though they are 
consistent within the large uncertainty from pDIS.  For GS-C 
and DSSV, which have a node in \dgx near the center of the 
sampled $x$ region, a cancelation between the positive and 
negative contribution in the wide $x$ distribution in each \pt 
bin leads to a small value of \tdg and thus small \all.  As 
these two parameterizations have significantly different 
$\dg^{[0,1]}$ values, it is clear that we are sensitive only 
to \tdg.

\begin{figure}[tbp]
\includegraphics[width=1.0\linewidth]{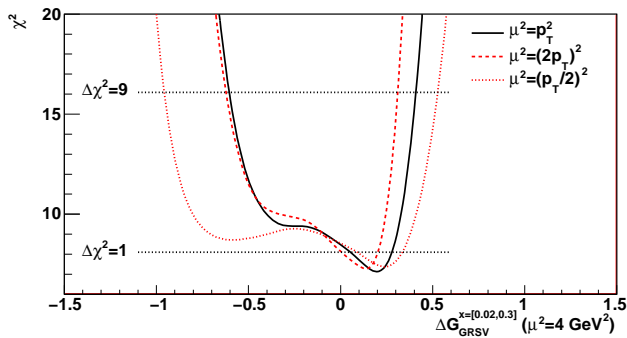}
\caption{\label{f:scale} (color online)
\chisq profile as a function of $\dg^{[0.02,0.3]}_{\rm GRSV}$ 
when the theoretical scale is set to $\mu=\pt$, $\pt/2$ and 
$2\pt$.}
\end{figure}

For each pDIS fit, by varying $\dg^{[0,1]}$ at the input 
scale while fixing the quark distributions and the shape 
of \dgx to the best fit values, a range of \all curves 
were generated.  Figure~\ref{f:model}b shows the resulting 
\chisq profiles.  While this approach is different from 
that in Fig.~\ref{f:data}b, resulting in a different 
\chisq profile shape for GRSV, the $\Delta\chisq$=1 and 9 
constraints are quite consistent.  The \tdg values at the 
\chisq minimum for each parameterization are between 0.1 
and 0.2, and are listed in Table~\ref{t:dgvalues}. At 
$\Delta\chisq$=9, the profiles are consistent for all 
parameterizations, independent of shape, indicating that 
our data are primarily sensitive to the size of \tdg.

The cross section for \pz production has been presented 
\cite{r:PHENIX:run5pi0all:Adare:2007dg} and compared with NLO 
pQCD expectations with the theoretical scales (factorization, 
fragmentation, and renormalization) in the calculation all set 
equal to $\mu=k\pt$ with $k=1$. The calculation agreed with 
our results within the sizable theoretical uncertainties in 
the choice of scale, which were estimated by varying $k$ up 
and down by a factor of two.  As we rely on NLO pQCD to 
extract \tdg from our measured \allpz, we must consider the 
effect of this uncertainty. Figure~\ref{f:scale} shows the 
change in the $\dg^{[0.02,0.3]}_{\rm GRSV}$ constraint when 
varying $k$ in the \all calculation in the GRSV 
parameterization.  The theoretical scale uncertainty for the 
constraint on positive values of $\dg^{[0.02,0.3]}_{\rm GRSV}$ 
is similar to that for varying the parameterization, while 
large uncertainty arises for negative values. 


We have presented results for \allpz from Run-6, which, 
combined with Run-5 results, at $\mu^2=4$ GeV$^2$ give
\begin{eqnarray}
\dg^{[0.02,0.3]}_{\rm GRSV}=&0.2&\pm0.1{\rm (stat)}\pm0.1{\rm (sys)} \nonumber \\
                                                          &&^{+0.0}_{-0.4}{\rm (shape)}\pm0.1{\rm (scale)}.
\end{eqnarray}
Using four parameterizations of \dgx, we find a shape 
independent constraint of $-0.7<\tdg<0.5$ at 
$\Delta\chisq$=9 ($\sim$3$\sigma$). The theoretical scale 
induced uncertainty is small for positive values of 
$\dg^{[0.02,0.3]}_{\rm GRSV}$, but is sizable for negative 
values.  Future measurements will be required to measure 
$\dgx$ for $x<0.02$ where large uncertainty remains 
\cite{r:DSSV:deFlorian:2008mr} and which may still 
contribute a significant amount of the proton spin. The 
quark spin contribution was well constrained by pDIS, and 
our result begins to significantly constrain the gluon 
spin contribution as well.


We thank the staff of the Collider-Accelerator and Physics 
Departments at BNL for their vital contributions. We 
acknowledge support from the Office of Nuclear Physics in DOE 
Office of Science, NSF, and a sponsored research grant from 
Renaissance Technologies (U.S.), MEXT and JSPS (Japan), CNPq 
and FAPESP (Brazil), NSFC (China), MSMT (Czech Republic), 
IN2P3/CNRS, and CEA (France), BMBF, DAAD, and AvH (Germany), 
OTKA (Hungary), DAE (India), ISF (Israel), KRF and KOSEF 
(Korea), MES, RAS, and FAAE (Russia), VR and KAW (Sweden), 
U.S. CRDF for the FSU, US-Hungary Fulbright, US-Israel BSF.



\hyphenation{Post-Script Sprin-ger}

\
\end{document}